\documentclass[twocolumn,amsmath,amssymb,prb,longbibliography]{revtex4-1}
\usepackage{multirow}
\usepackage{array}
\usepackage{amsmath}
\usepackage{graphicx}
\usepackage{dcolumn}
\usepackage{bm}
\usepackage{verbatim}
\DeclareMathAlphabet \mathbfcal{OMS}{cmsy}{b}{n}
\begin{document}



\title{Ultrafast topological phenomena in gapped graphene}

\author{S. Azar Oliaei Motlagh}
\author{Vadym Apalkov}
\author{Mark I. Stockman}
\affiliation{Center for Nano-Optics (CeNO) and
Department of Physics and Astronomy, Georgia State
University, Atlanta, Georgia 30303, USA
}

\date{\today}
\begin{abstract}
In the model of a gapped graphene, we have shown how the recently predicted topological resonances are solely related to the presence of an energy band gap at the $K$ and $K^\prime$ points of the Brillouin zone. In the field of a strong single-oscillation chiral (circularly-polarized) optical pulse, the topological resonance causes the valley-selective population of the conduction band. This population distribution represents a chiral texture in the reciprocal space that is structured with respect to the pulse separatrix as has earlier been predicted for transition metal dichalcogenides. As the band gap is switched off, this chirality gradually disappears replaced by a achiral distribution characteristic of graphene.

\end{abstract}
\maketitle
\section{Introduction} 

Two dimensional (2D) materials with hexagonal symmetry \cite{Novoselov_et_al_Science_2016_2D_Materials_and_Heterostructures} -- graphene, silicene, transition metal dichalcogenides (TMDC's),   hexagonal boron nitride (h-BN), etc. -- possess nontrivial topological properties in the reciprocal space \cite{Xiao_Niu_RevModPhys.82_2010_Berry_Phase_in_Electronic_Properties}. We aim at study of nonlinear behavior of such materials in strong ultrafast (one or a few optical oscillations) laser pulse fields. This behavior was predicted to be fundamentally different for graphene and TMDC's \cite{Stockman_et_al_PRB_2016_Graphene_in_Ultrafast_Field, Stockman_PhysRevB.92_2015_Ultrafast_Control_Symmetry, Stockman_et_al_PhysRevB.93.155434_Graphene_Circular_Interferometry, Stockman_et_al_PhysRevB.98_2018_Rapid_Communication_Topological_Resonances}. 

In graphene, linearly-polarized pulses cause appearance of interference fringes in the reciprocal space, which are due to transitions in the vicinity of the $K$ (or, $K^\prime$) point that occur twice per cycle; the corresponding transition amplitudes interfere causing the fringes \cite{Stockman_et_al_PRB_2016_Graphene_in_Ultrafast_Field}. For a circular polarized pulses, there is no effect of pulse handedness for a single optical oscillation. For two or more optical oscillations there is weak preferential population of one of the $K$ or $K^\prime$ valley and and characteristic forks of the electron interferogram indicating effect of the Berry phase. The electron distribution for both linear and circular-polarized pulses are asymmetric, which causes currents that have recently been observed experimentally \cite{Higuchi_Hommelhoff_et_al_Nature_2017_Currents_in_Graphene}.

In a contrast, for TMDC's there is a strong preferential population of the valley that corresponds by its chirality to the handedness of the circularly polarized excitation pulse (called the valley polarization). There is also texturing of the reciprocal space with respect to the separatrix [see below Eq.\ (\ref{separatrix}) for definition]. Namely, the preferentially populated valley is populated outside of the separatrix while the low-population valley is populated inside the separatrix. These phenomena, which we called the topological resonance  \cite{Stockman_et_al_PhysRevB.98_2018_Rapid_Communication_Topological_Resonances}, are due to the interference of the topological (Berry) phase and the dynamic phase of the polarization oscillation.

In this article, we explore a model of gapped graphene \cite{Kjeld_et_al_PhysRevB.79.113406_2009_Gapped_Graphene_Optical_Response} where the center symmetry is removed by introducing sublattice-specific oncite energies $\pm\Delta$ -- see below Eq.\ (\ref{H0}); this opens up a band gap of $2\Delta$ at the $K$ and $K^\prime$ points. We show that, as the band gap increases,  the distribution of the carriers in the reciprocal space gradually changes from that characteristic of graphene with a low valley polarization to a dramatically different texture characteristic of TMDC's with a high valley polarization. Note that experimentally the band gap in graphene can be open, in particular, by growing it on a SiC substrate \cite{Ajayan_et_al_Review-JNN_2011_Band_Opening_in_Graphene, Conrad_et_al_PhysRevLett.115_2015_Gapped_Graphene_on_SiC}.
   
Graphene, a two dimensional (2D) layer of carbon atoms with a honeycomb symmetry, possesses unique physical properties: gapless Dirac-fermion spectrum at the $K$ and $K^\prime$ points, nonzero Berry curvature concentrated at these Dirac points corresponding to the $\pm\pi$ Berry phase, tunable carrier density and plasmonic properties, unusual magnetic properties including the quantum Hall effect at room temperatures, etc. \cite{Kane_Mele_PhysRevLett.95_Spin_Hall_Effect_in_Graphene, Kim_et_al_2005_nature04235_Quantum_Hall_Effect_in_Graphene, Novoselov_Geim_et_al_nature04233_2D_Electrons_in_Graphene, Novoselov_et_al_Nature_Materials_2007_Rise_of_Graphene, Electronic_properties_graphene_RMP_2009, Grigorenko_et_al_Nat_Phot_2012_Graphene_Plasmonics, Novoselov_et_al_Nature_2012_Graphene_Review, Geim_et_al_Physrevlett.111.166601_2013_Giant_Magnetodrag,  Wang_et_al_PhysRevLett.114_2015_Graphene_Ferromagnet, Novoselov_et_al_Science_2016_2D_Materials_and_Heterostructures}. 

To elucidate high-field ultrafast behavior of graphene, we have theoretically studied its behavior for linear-polarized and chiral (circularly-polarized) few-oscillation optical pulses, which cause population transfer from the valence band (VB) to the conduction band (CB). We found that linearly-polarized pulses caused appearance of interference fringes  in the CB population to the quantum interference of two passages of electrons by the Dirac points where VB$\to$CB transitions occurred \cite{Stockman_et_al_PRB_2016_Graphene_in_Ultrafast_Field}. This quantum interference also caused field-induced currents and total charge transfer per pulse (optical rectification) predicted in Ref.\ \onlinecite{Stockman_et_al_PRB_2016_Graphene_in_Ultrafast_Field}. However, as expected, the linear pulse was ``blind'' to the valley chirality: the CB population distribution in the $K$ and $K^\prime$ valleys were exactly the same as protected by the time reversal ($\cal T$) symmetry.

The chiral single-oscillation optical pulses also did not produce 
any significant valley-specific CB population or interference fringes, which was explained by the fact that an electron experiences only a single passage in the vicinity of a Dirac point. For a pulse with a few (two or more) optical oscillations, the quantum pathways of the passages by a Dirac point for different optical cycles would interfere causing the appearance of pronounced chiral interference patterns \cite{Stockman_et_al_PhysRevB.93.155434_Graphene_Circular_Interferometry}. These were different for the $K$ and $K^\prime$ valleys depending on whether the pulse chirality is the same or opposite to that of the corresponding valley. These chiral structures contained characteristic ``forks'' revealing vortices of the Berry connection. The resulting electron distributions in the reciprocal space were asymmetric, which obviously would lead to electron currents. Experimentally, such electric currents in graphene induced by both linearly- and circularly-polarized optical pulses were observed in Ref.\ \onlinecite{Higuchi_Hommelhoff_et_al_Nature_2017_Currents_in_Graphene}.

We have also studied another class of hexagonal-symmetry 2D systems -- transition metal dichalcogenides (TMDC's) such as MoS$_2$ and WS$_2$ \cite{Stockman_et_al_PhysRevB.98_2018_Rapid_Communication_Topological_Resonances}. In a dramatic contrast to graphene, the TMDC's showed very strong valley selectivity (preference to valley chirality) even for a single-oscillation circularly-polarized optical pulse where the valley polarization was $\gtrsim 80\%$. There are two differences with respect to graphene: (i) the TMDC's  have a band gap (these TMDC's are direct band-gap semiconductors) and (ii) The TMDC's have a significant spin-orbit interaction splitting both the CB's and the VB's. We have shown that the TMDC's in the circularly-polarized pulse field exhibit a new type of resonance -- the topological resonance -- where the reciprocal space is textured with respect to a topologically-defined curve called separatrix. For a given pulse chirality, one valley has a high CB population outside the separatrix while the other valley has a low CB population inside the separatrix. Such a structuring is completely absent in graphene.

In this article, we employ a model of gapped graphene  \cite{Kjeld_et_al_PhysRevB.79.113406_2009_Gapped_Graphene_Optical_Response} where the band gap can be opened and continuously tuned controlled by an on-site energy $\Delta$ parameter. This material is subjected to a strong-field single-oscillation optical pulse. We aim to investigate the transition from an almost achiral strong-field population of the graphene CB with no signs of the topological resonances and a  low valley polarization to a highly-chiral and valley-selective population of the CB's in the TMDC's, which does exhibit a pronounced topological resonance and a high valley polarization. 

The gapped graphene is described by a two-band model for direct band gap 2D semiconductors with the $K$ and the $K^\prime$ valleys mimicking the bilayer graphene, h-BN, or TMDC's. We use a two band tight-binding Hamiltonian of graphene where we additionally introduce an adjustable gap through different on-site energies, $\Delta$ and $-\Delta$, for the two sublattices, A and B. This difference of the on-site energies causes the breakdown of the inversion symmetry and opens up the  band gaps of $2\Delta$ at the  $K$ and the $K^\prime$ points whose equality is protected by the $\cal T$-inversion symmetry.

\section{MODEL AND MAIN EQUATIONS}
\label{Model_and_Equations}

\begin{figure}
\begin{center}\includegraphics[width=0.47\textwidth]{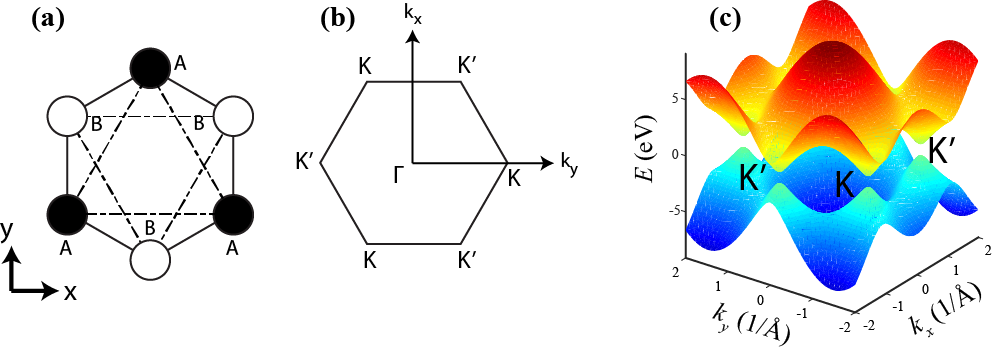}\end{center}
  \caption{(Color online) (a) Hexagonal lattice structure of graphene with sublattices A and B. (b) The first Brillouin zone of the reciprocal lattice of graphene with two valleys $K$ and $K^\prime$. (c) Energy dispersion is shown as a function of crystal momentum for gapped graphene (1 eV)}
  \label{fig:Energy}
\end{figure}

The electron-collision relaxation times in graphene and 2D materials are on the order or significantly longer than 10 fs. \cite{Hwang_Das_Sarma_PRB_2008_Graphene_Relaxation_Time, Breusing_et_al_Ultrafast-nonequilibrium-carrier-dynamics_PRB_2011, theory_absorption_ultrafast_kinetics_graphene_PRB_2011, Ultrafast_collinear_scattering_graphene_nat_comm_2013, Gierz_Snapshots-non-equilibrium-Dirac_Nat-Material_2013, Nonequilibrium_dynamics_photoexcited_electrons_graphene_PRB_2013}. Therefore, for an ultrashort optical pulse with the duration of less than 10 fs, we assume that the electron dynamics in the field of the pulse is coherent and the electron collision effects are negligible. With the described assumption, the electron dynamics is described by the time-dependent Schr\"odinger equation (TDSE), which has the following form 
\begin{equation}
i\hbar \frac{{d\Psi }}{{dt}} = { H(t)} \Psi  
\label{Sch}
\end{equation}
with Hamiltonian
\begin{equation}
{ H}(t) = { H}_0 - e{\bf{F}}(t){\bf{r}},
\label{Ht}
\end{equation}  
where $\mathbf F(t)$ is the  pulse's electric field, $e$ is electron charge, and $H_0$ is the nearest neighbor tight binding Hamiltonian for gapped graphene,
\begin{eqnarray}
H_0=\left( {\begin{array}{cc}
   \Delta & \gamma f(\mathbf k) \\
   \gamma f^\ast(\mathbf k) & -\Delta \\
  \end{array} } \right) ,
\label{H0}
\end{eqnarray}
2$\Delta$ is the aforementioned finite gap between the CB and the VB,
$\gamma= -3.03$ is hopping integral, and
\begin{equation}
f(\mathbf k)=\exp\Big(i\frac{ak_y}{\sqrt{3}}\Big )+2\exp\Big(-i\frac{ak_y}{2\sqrt{3}}\Big )\cos{\Big(\frac{ak_x}{2}\Big )},
\end{equation}
where $a=2.46~\mathrm{\AA}$ is lattice constant. The energies of CB and VB can be found from the above Hamiltonian, $H_0$, as the following expressions
\begin{eqnarray}
E_{c}(\mathbf k)&=&+\sqrt{\gamma ^2\left |{f(\mathbf k)}\right |^2+\Delta ^2}~~,
\nonumber \\
E_{v}(\mathbf k)&=& -\sqrt{\gamma ^2\left |{f(\mathbf k)}\right |^2+\Delta ^2}~~,
\label{Energy}
\end{eqnarray}
where $c$ and $v$ stand for the CB and VB, respectively. This energy dispersion is shown in Fig. \ref{fig:Energy}(c).

Below we assume that the VB is fully occupied and the CB is empty.
We will be applying intense fields with amplitude $F_0\gtrsim0.1~\mathrm{V/\AA}$. At such intensities, the number of photons, $N_p$, per pulse within the minimum coherence area of $\sim\lambda^2$, where $\lambda\sim 1~\mathrm{\mu m}$ is wavelength,  
\begin{equation}
N_p\sim \frac{c \tau_p \lambda^2 F_0^2}{4\pi \hbar\bar\omega}\sim 5\times 10^7~,
\end{equation}
where $c$ is speed of light; we assume realistic parameters: $\tau_p\sim3~\mathrm{fs}$ is the pulse duration, and $\hbar\bar\omega\sim 1~\mathrm{eV}$ is the mean photon energy. With such a large photon number involved, it is legitimate to describe $\mathbf F(t)$ as a classical electric field keeping the quantum-mechanical description for the solid. This is a usual semi-classical approach in high-field optics -- see, e.g., Refs.\ \onlinecite{Corkum_Krausz_Nature_Physics_2007_Attosecond_Science, Krausz_Ivanov_RevModPhys.81.163_2009_Attosecond_Review, Krausz_Stockman_Nature_Photonics_2014_Attosecond_Review}. Note that quantized optical fields are used for much lower intensities 
\cite{Kibis_PhysRevB.81_2010_MI_Transition_in_Graphene_by_CP_Photons, Kristinsson_Sci_Rep-2016-Control_of_electronic_transport_in_graphene}. Such a full quantum mechanical approach is not needed for the fields of the amplitude we consider. We solve the Schr\"odinger equation in the truncated basis of Houston functions (\ref{Houston}) numerically without further approximations. Our pulse is just a single optical oscillation; therefore field $\mathbf F(t)$ is not periodic, and its effect cannot be described as band gap modification as in Refs.\ \onlinecite{Kibis_PhysRevB.81_2010_MI_Transition_in_Graphene_by_CP_Photons, Classen_et_al_Nat_Commun_2016_All_Optical_Chiral_Edge_Modes}. However, the dynamic Stark effect and other field-dressing effects during the pulse are indeed fully taken into account by our solution.

In solids, the applied electric field generates both the intraband and interband electron dynamics. The intraband dynamics is determined by the Bloch acceleration theorem \cite{Bloch_Z_Phys_1929_Functions_Oscillations_in_Crystals} for time evolution of the crystal momentum, $\mathbf k$, 
\begin{equation}
\hbar \frac{{d{\bf{k}}}}{{dt}} = e{\bf{F}}(t).
\label{acceleration}
\end{equation}
From this, for an electron with an initial crystal momentum ${\bf q}$,   time-dependent crystal momentum ${\mathbf k }({\mathbf q},t )$ is expressed as
\begin{equation}
{{\bf{k}}}({\bf{q}},t) = {\bf{q}} + \frac{e}{\hbar }\int_{ - \infty }^t {{\bf{F}}({t^\prime})d{t^\prime}}. 
\label{kvst}
\end{equation}

Related to Bloch trajectories (\ref{kvst}), we also define 
the separatrix as a set of initial points $\mathbf q$ for which electron trajectories pass precisely through the corresponding $K$ or $K^\prime$ points \cite{Stockman_et_al_PhysRevB.93.155434_Graphene_Circular_Interferometry}. Its parametric equation is 
\begin{equation}
\mathbf q(t)=\mathbf K-\mathbf k(0,t), \mathrm{~~or~~}  \mathbf q(t)=\mathbf K^\prime-\mathbf k(0,t),
\label{separatrix}
\end{equation}
 where $t\in (-\infty,\infty)$ is a parameter. 

The corresponding wave functions, which are solutions of Schr\"odinger equation (\ref{Sch}) within a 
single band $\alpha$, i.e., without interband coupling, are the well-known Houston functions \cite{Houston_PR_1940_Electron_Acceleration_in_Lattice},
\begin{equation}
\Phi _{\alpha {\bf{q}}}^{(H)}({\bf{r}},t) = \Psi _{{{\bf{k}}}({\bf{q}},t)}^{(\alpha )}({\bf{r}}){e^{ - \frac{i}{\hbar }\int_{ - \infty }^t {d{t_1}{E_\alpha }[{{\bf{k}}}({\bf{q}},{t_1})]} }}~,
\label{Houston}
\end{equation}
where $\alpha=v,c$ for the VB and CB, correspondingly, and $ \mathrm{\Psi^{(\alpha)}_{{\mathbf k}}} $ are Bloch-band eigenfunctions in the absence of the pulse field, and $E_\alpha(\mathbf k)$ are the corresponding eigenenergies.

The interband electron dynamics is determined by the solution of the TDSE (\ref{Sch}).  Such a solution can be expanded in the basis of the Houston functions $\Phi^{(H)}_{\alpha {\bf q}}({\bf r},t)$,
\begin{equation}
\Psi_{\bf q} ({\bf r},t)=\sum_{\alpha=c,v}\beta_{\alpha{\bf q}}(t) \Phi^{(H)}_{\alpha {\bf q}}({\bf r},t),
\end{equation}
where 
$\beta_{\alpha{\bf q}}(t)$ are expansion coefficients.

Let us introduce the following quantities
\begin{eqnarray}
&&\mathbfcal D^{cv}(\mathbf q,t)=
\mathbfcal A^{cv}[\mathbf k (\mathbf q,t)]\exp\left(i\phi^\mathrm{(d)}_{cv}(\mathbf q,t)\right),
 \label{Q}
\\
&&\phi^\mathrm{(d)}_{cv}(\mathbf q,t)=
\nonumber
\\
 && \frac{1}{\hbar} \int_{-\infty}^t dt^\prime \left(E_c[\mathbf k (\mathbf q,t^\prime)]-E_{v}[\mathbf k (\mathbf q,t^\prime)]\right),
 \label{phi}
 \\ 
&&{\mathbfcal{A}}^{cv}({\mathbf q})=
\left\langle \Psi^{(c)}_\mathbf q  |   i\frac{\partial}{\partial\mathbf q}|\Psi^{(v)}_\mathbf q   \right\rangle .
\label{D}
\end{eqnarray} 
Here $ \mathrm{\Psi^{(v)}_{{\mathbf q}}} $ and $ \mathrm{\Psi^{(c)}_{{\mathbf q}}} $ are periodic Bloch functions, i.e., eigenfunctions of the Hamiltonian without an optical field;   ${\mathbfcal A}^{cv}(\mathbf q)$ is  a matrix element of the well-known non-Abelian Berry connection \cite{Wiczek_Zee_PhysRevLett.52_1984_Nonabelian_Berry_Phase, Xiao_Niu_RevModPhys.82_2010_Berry_Phase_in_Electronic_Properties, Yang_Liu_PhysRevB.90_2014_Non-Abelian_Berry_Curvature_and_Nonlinear_Optics}, and $\phi^{\mathrm{(d)}}_{cv}(\mathbf q,t)$ is the dynamic phase; the trajectory in the reciprocal space, $\mathbf k(\mathbf q, t)$, is given by the Bloch theorem (\ref{kvst}). Note that the interband dipole matrix element, which determines optical transitions between the VB and the CB at crystal momentum $\mathbf q$, is $\mathbf D^{cv}(\mathbf q)=e \mathbfcal{A}^{cv}(\mathbf q)$.

The non-Abelian Berry connection matrix elements can be found analytically as
\begin{eqnarray}
\mathcal{A}_{x}^{cv}(\mathbf k)&=&\mathcal N\Bigg(\frac{-a}{2|f(\mathbf k)|^2}\Bigg)\Bigg( \sin\frac{ak_x}{2}\sin\frac{a\sqrt{3}k_y}{2}
\nonumber\\
&&+i \frac{\Delta}{E_c}\Bigg(\cos \frac{a\sqrt{3}k_y}{2}\sin \frac{ak_x}{2}+\sin{ak_x}\Bigg)\Bigg)
\nonumber \\
 \label{Ax}
\\
\mathcal{A}_{y}^{cv}(\mathbf k)&=&\mathcal N\Bigg(\frac{a}{2\sqrt{3}|f(\mathbf k)|^2}\Bigg)\Bigg( -1-\cos\frac{a\sqrt{3}k_y}{2}\cos\frac{ak_x}{2}
\nonumber\\
&&+2\cos ^2 \frac{ak_x}{2}-i \frac{3\Delta}{E_c}\sin \frac{a\sqrt{3}k_y}{2}\cos \frac{ak_x}{2}\Bigg)
\nonumber \\
\label{Ay}
\end{eqnarray}
where
\begin{equation}
\mathcal N=\frac{\left|\gamma f(\mathbf k)\right|}{\sqrt{\Delta^2+\left|\gamma f(\mathbf k)\right|^2}}~.
\end{equation}

In these terms, we introduce Schr\"odinger equation in the interaction representation in the adiabatic basis of the Houston functions as
\begin{equation}
i\hbar\frac{\partial B_\mathbf q(t)}{\partial t}= H^\prime(\mathbf q,t){B_\mathbf q}(t)~,
\label{Schrodinger}
\end{equation}
where wave function (vector of state) $B_q(t)$ and Hamiltonian $ H^\prime(\mathbf q,t)$ are defined as 
\begin{eqnarray}
B_\mathbf q(t)&=&\begin{bmatrix}\beta_{c\mathbf q}(t)\\ \beta_{v\mathbf q}(t)\\ \end{bmatrix}~,\\ 
H^\prime(\mathbf q,t)&=&-e\mathbf F(t)\mathbfcal{\hat A}(\mathbf q,t)~,\\
\mathbfcal{\hat A}(\mathbf q,t)&=&\begin{bmatrix}0&\mathbfcal D^{cv}(\mathbf q,t)\\
\mathbfcal D^{vc}(\mathbf q,t)&0\\
\end{bmatrix}~.
\end{eqnarray}

 Schr\"odinger equation (\ref{Schrodinger}) defines a solution for dynamics of the system, whose accuracy is limited by the size of the basis set (i.e., truncation of the Hilbert space of the crystal). In particular, it contains such phenomenon as band gap opening in the field of a circularly-polarized pulse. A formal general solution of this equation can be presented in terms of the evolution operator, $\hat S(\mathbf q,t)$, as
 \begin{eqnarray}
 B_\mathbf q (t)&=&\hat S(\mathbf q,t)B_\mathbf q (-\infty)~,\\
 \hat S(\mathbf q,t)&=&\hat T \exp\left[i\int_{-\infty}^t \mathbfcal{\hat A}(\mathbf q,t^\prime)d\mathbf k(t^\prime)\right]~,
 \label{S}
 \end{eqnarray}
where $\hat T$ is the well-known time-ordering operator \cite{Abrikosov_Gorkov_Dzialoshinskii_1975_Methods_of_Quantum_Field_Theory}, and the integral is affected along the Bloch trajectory [Eq.\ (\ref{kvst})]: $d\mathbf k(t)=\frac{e}{\hbar}\mathbf F(t)dt$.

\begin{figure}
\begin{center}\includegraphics[width=0.47\textwidth]{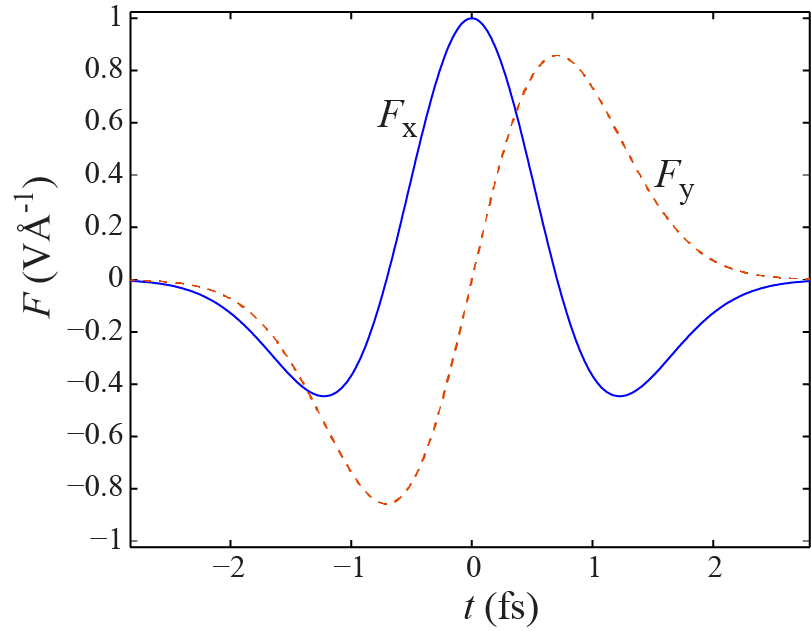}\end{center}
  \caption{(Color online) $F_x$ and $Fy$ components of right handed circularly polarized field.}
  \label{fig:field}
\end{figure}%

\begin{figure}
\begin{center}\includegraphics[width=0.47\textwidth]{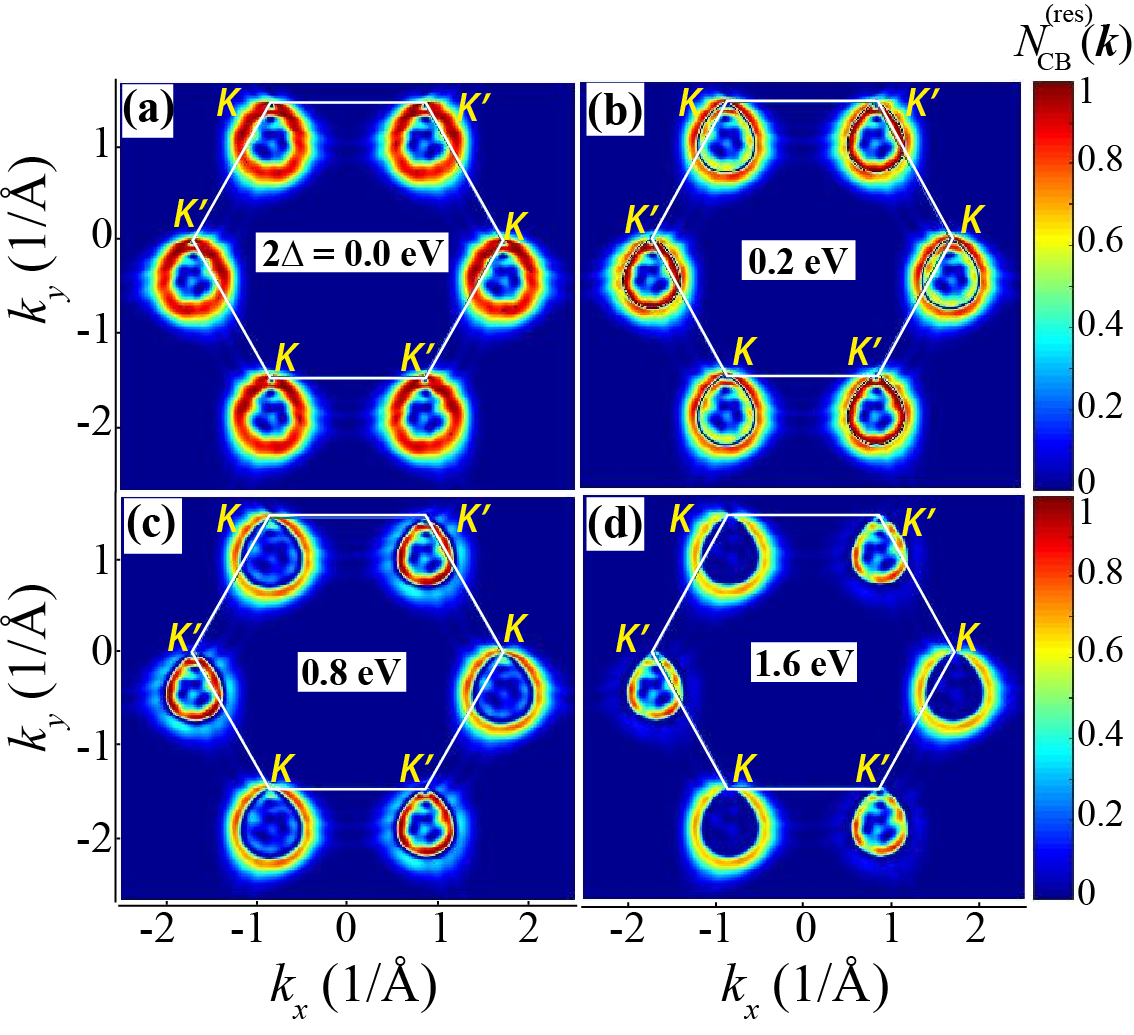}\end{center}
  \caption{(Color online) Residual CB population $N\mathrm{^{(res)}_\mathrm{CB}}(\mathbf{k})$ for graphene with adjustable bandgap in the extended zone picture. The white solid line shows the boundary of the first Brillouin zone with $K, K^\prime$-points indicated. The amplitude of the optical field with the right handed polarization is 0.5 $\mathrm{V\AA^{-1}}$.  The band gap is 0 (a), 0.2 eV (b), 
0.8 eV (c), and 1.6 eV (c).}
  \label{fig:1RC_F0=0p5VpA_gapped_graphene}
\end{figure}%

\begin{figure}
\begin{center}\includegraphics[width=0.47\textwidth]{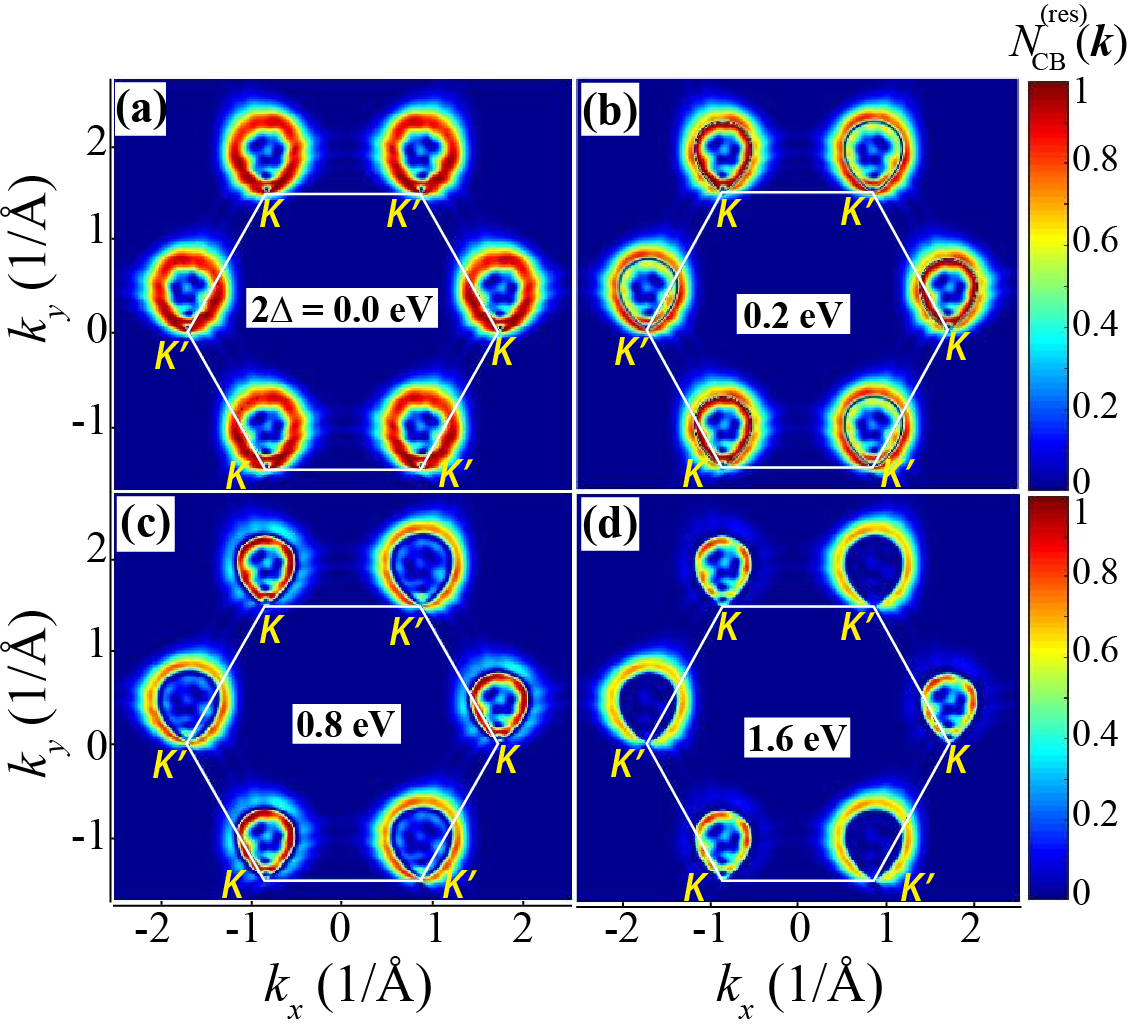}\end{center}
  \caption{(Color online) Residual CB population $N\mathrm{^{(res)}_\mathrm{CB}}(\mathbf{k})$ for graphene with adjustable bandgap in the extended zone picture. The white solid line shows the boundary of the first Brillouin zone with $K, K^\prime$-points indicated. The amplitude of the optical field with the left handed polarization is 0.5 $\mathrm{V\AA^{-1}}$.  The band gap is 0 (a), 0.2 eV (b), 
0.8 eV (c), and 1.6 eV (c).}
  \label{fig:1CT_F0=0p5VpA_gapped_graphene}
\end{figure}%

\begin{figure}
\begin{center}\includegraphics[width=0.47\textwidth]{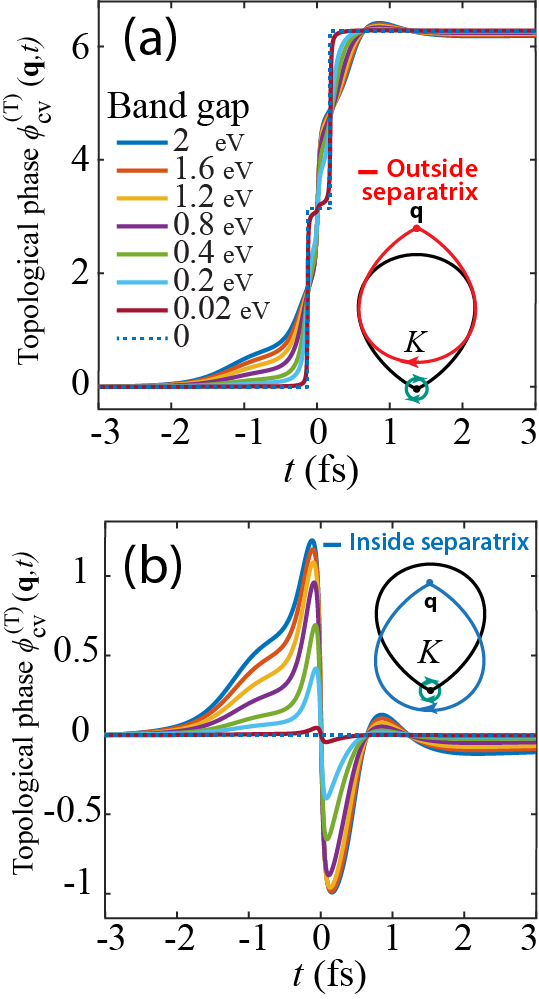}\end{center}
  \caption{(Color online) 
Topological phase $\phi_\mathrm{cv}^\mathrm{(T)}(\mathbf q,t)$
as a function of time for gapped graphene in 
the field of left-handed optical pulse with the amplitude of 0.5 V$\mathrm\AA^{-1}$. The topological phase is calculated along 
the electron trajectory in the reciprocal space for (a) initial 
$\mathbf q $ point outside of the sepratrix and (b) initial 
$\mathbf q $ point inside of the sepratrix.  
Inset: solid black line illustrates the separatrix for $K^\prime$ valley,
while the red line in panel (a) and the blue line in panel (b) 
show the corresponding electron trajectories. 
}
  \label{fig:Phase_F_dot_D_CT_F0_5}
\end{figure}%

\begin{figure}
\begin{center}\includegraphics[width=0.47\textwidth]{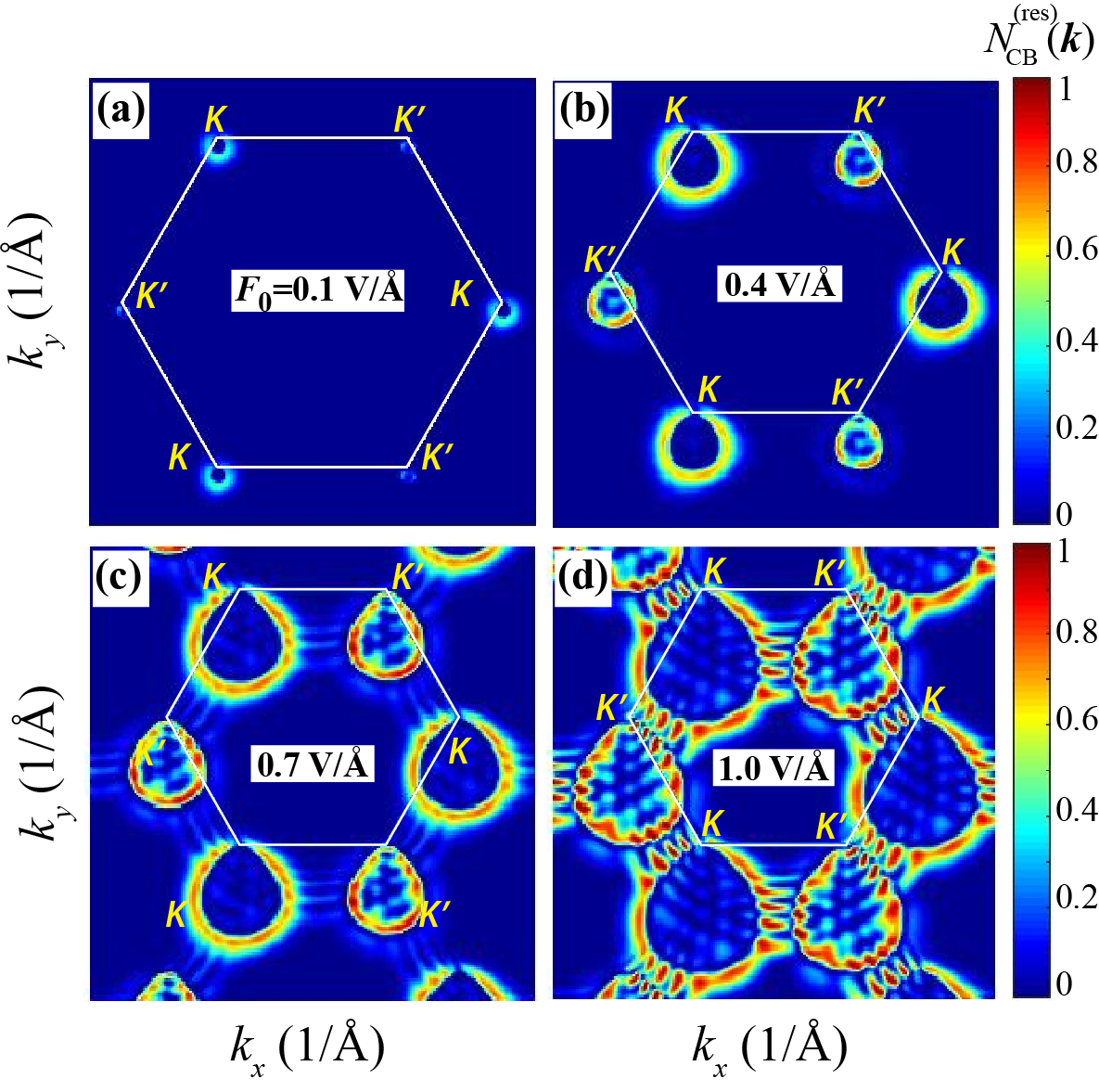}\end{center}
  \caption{(Color online) Residual CB population $N\mathrm{^{(res)}_\mathrm{CB}}(\mathbf{k})$ for graphene with 1.6 eV bandgap in the extended zone picture. The white solid line shows the first Brillouin zone boundary with $K, K^\prime$-points indicated. The amplitude of the optical field with right handed polarization is 0.5 $\mathrm{V\AA^{-1}}$.  Residual population, $N\mathrm{^{(res)}_\mathrm{CB\uparrow}}(\mathbf{k})$, for (a) field amplitude is equal to 0.1 $\mathrm{V\AA^{-1}}$, (b) field amplitude is equal to 0.4 $\mathrm{V\AA^{-1}}$, (c) field amplitude is equal to 0.7 $\mathrm{V\AA^{-1}}$, and (d) field amplitude is equal to 1 $\mathrm{V\AA^{-1}}$. }
  \label{fig:1RC_gap=1p6eV_gapped_graphene}
\end{figure}%

The total charge current, ${\bf J}(t) = \left\{J_x(t),J_y(t)\right\}$, generated during the pulse is the summation of the interband and intraband currents, and it is determined by the following expression
\begin{equation}
{J_j}(t) = \frac{e}{{{a^2}}}\sum\limits_{\mathbf{q}} {\sum\limits_{{\alpha _1,\alpha_2} = v ,c}  {\beta^\ast _{{\alpha _1}{\mathbf{q}}}(t)V\;_j^{{\alpha _1}{\alpha _2}}(\mathbf k\left(\mathbf q, t)\right){\beta _{{\alpha _2}{\bf{q}}}}(t)} }  ,
\label{J}
\end{equation}
where $j = x, y$, a =2.46 $\mathrm{\AA}$ is lattice constant, and $V_j^{{\alpha _1}{\alpha _2}}(\mathbf k)$ are matrix elements of the velocity operator,
\begin{equation}
\hat{V}_j = \frac{1}{\hbar }\frac{{\partial {H_0}}}{{\partial {k_j}}}.
\end{equation}
For the known eigenstates, the intraband velocities are calculated as following
\begin{eqnarray}
V_x^{\mathrm{ cc}}(\mathbf k)&=&-V_x^{\mathrm{vv}}(\mathbf k)=\frac{-a\gamma ^2}{\hbar{\sqrt{|\gamma{f(\mathbf k)}|^2+\Delta^2}}}
\nonumber
\\
&\times&\sin{\frac{ak_x}{2}}\Big(\cos{\frac{\sqrt{3}ak_y}{2}}
+2\cos{\frac{ak_x}{2}}\Big)\\
V_y^{\mathrm{ cc}}(\mathbf k)&=&-V_y^{\mathrm{vv}}(\mathbf k)=\frac{-\sqrt{3}a\gamma ^2}{\hbar{\sqrt{|\gamma{f(\mathbf k)}|^2+\Delta^2}}}
\nonumber
\\
&\times&\sin{\frac{\sqrt{3}ak_y}{2}}\cos{\frac{ak_x}{2} }.
\end{eqnarray}

The interband velocities can be expressed in terms of the non-Abelian Berry connection matrix elements as the following
\begin{eqnarray}
V_x^{\mathrm{cv}}=\frac{i}{\hbar}{\mathcal A}^{cv}_x(E_c-E_v)~,
\nonumber 
\\ 
V_y^{\mathrm{cv}}=\frac{i}{\hbar}{\mathcal A}^{cv}_y(E_c-E_v)~.
\end{eqnarray}

\section{Results and discussion}
\subsection{Circularly polarized pulse}
Consider the ultrafast valley polarization induced by a circularly polarized pulse in the gapped graphene. We apply an ultrafast circularly-polarized optical pulse $\mathbf F$=($F_x$, $F_y$), which is parametrized as the following 
\begin{eqnarray}
F_x&=&F_0(1-2u^2)e^{-u^2}
\label{Fx}
\\
F_y&=&\pm 2F_0ue^{-u^2}
\label{Fy}
\end{eqnarray}
Here, $\pm$ determines the handedness: the upper sign is for the
right-handed circular polarization, and the lower is for the left-handed circular polarization, which is $\cal T$-reversed with respect to the first one; $F_0$ is the amplitude of the optical oscillation, and $u=t/\tau$ , where $\tau$ is a characteristic half duration of the optical oscillation (in calculations, we choose $\tau$ = 1 fs). The $x$ and $y$ components of $\mathbf F(t)$ [Eqs.\ (\ref{Fx}) and (\ref{Fy})] for the right-handed circularly polarized pulse are displayed in Fig.\ \ref{fig:field}.

By using the theory described above in Sec.\ \ref{Model_and_Equations}, we solve TDSE (\ref{Schrodinger}) numerically with the initial condition ($\beta_{c\mathbf q},\beta_{v\mathbf q})=(0,1)$. For an electron, which is initially in the VB, the applied optical field causes transitions into the CB and results in a finite CB population.

For an applied single cycle of a right-handed circularly-polarized pulse with the amplitude of 0.5 $\mathrm{V\AA^{-1}}$, the CB population after the pulse ends, known as residual CB population, $N\mathrm{^{(res)}_\mathrm{CB}}(\mathbf{q})=|\beta_{c\mathbf q}(t=\infty)|^2$,  is shown in Figs \ref{fig:1RC_F0=0p5VpA_gapped_graphene}(a)-\ref{fig:1RC_F0=0p5VpA_gapped_graphene}(d) for different values of the band gap. In the case of graphene, when the band gap is zero  [Fig. \ref{fig:1RC_F0=0p5VpA_gapped_graphene}(a)], the optical pulse populates the CB along the corresponding separatrix but does not produce any appreciable interference fringes or hot spots. 

For the case of graphene (the band gap is zero), the distributions of the residual CB population in the $K$ and $K^\prime$ valleys in Fig.\ \ref{fig:1RC_F0=0p5VpA_gapped_graphene}(a) are very close to each other. However, there are some small differences, especially visible inside the separatrix, which are mirror images of each other due to the reflection (${\cal P}_y$) symmetry of the lattice.

In the reciprocal space near the $K$ valley, with an increase of the band gap,  the area inside the separatrix get less populated in comparison to the area outside of separatrix [Figs.\ \ref{fig:1RC_F0=0p5VpA_gapped_graphene}(b)-(d)]. The opposite happens for the $K^\prime$ valley where the majority of the population is inside of separatrix. For the  $\cal T$-reversed (left handed) pulse, the distributions shown in Fig.\ \ref{fig:1CT_F0=0p5VpA_gapped_graphene} are  $\cal T$ reversed (or, center-reflected) images of the distributions in  Fig.\ \ref{fig:1RC_F0=0p5VpA_gapped_graphene}; in particular,  the $K$ and $K^\prime$ valleys are exchanged places. 

As we can see from comparison of the cases of a different band gap [different panels in Figs.\ \ref{fig:1RC_F0=0p5VpA_gapped_graphene}(a)-(d) or in Figs.\ \ref{fig:1CT_F0=0p5VpA_gapped_graphene}(a)-(d)], we conclude that with an increase of the band gap, the $K$ and $K^\prime$ valleys become increasingly populated differently (valley polarization); simultaneously asymmetry of the population with respect to  the separatrix appears: the major (dominating) population occurs outside of the separatrix while the minor population is inside.

Note that the separatrix is a topological object: it divides the reciprocal space into two distinct regions: any pulse-field-induced electron Bloch trajectory, which originates inside the separatrix, encircles the $K$ (or, $K^\prime$) point that is the center of the topological (Berry) curvature. To the opposite, a Bloch trajectory, which originates from the outside of the separatrix, does not encircle the $K$ (or, $K^\prime$) point. This difference causes an effect of the topological resonance \cite{Stockman_et_al_PhysRevB.98_2018_Rapid_Communication_Topological_Resonances}, which we briefly explain below. 

The fundamental evolution operator (\ref{S}) can be rewritten in the form
\begin{equation}
\hat S(\mathbf q,t)=\hat T\exp{\left[i\int_{-\infty}^t\mathcal{\hat A}_\Vert(\mathbf q,t^\prime)dk(t)\right]}~,
\label{S1}
\end{equation}
where a longitudinal component of the non-Abelian Berry connection is defined as $\mathcal{\hat A}_\Vert(\mathbf q, t)=\mathbfcal{\hat A}(\mathbf q,t)\mathbf F(t)/F(t)$, and $dk(t)=\frac{e}{\hbar}F(t) dt$. Explicitly, matrix $\mathcal{\hat A}_\Vert(\mathbf q,t)$ has the form
\begin{equation}
\mathcal{\hat A}_\Vert(\mathbf q,t)=\begin{bmatrix}0&\mathcal{D}^\mathrm{(cv)}_\Vert(\mathbf q,t)
\\
\mathcal{D}^{\mathrm{(cv)}\ast}_\Vert(\mathbf q,t)&0\end{bmatrix}~,
\label{Aphi}
\end{equation}
where
\begin{equation}
\mathcal{D}^\mathrm{(cv)}_\Vert(\mathbf q,t)=\left\vert\mathcal{  A}^\mathrm{(cv)}_\Vert(\mathbf k(\mathbf q,t)\right\vert \exp{\left[i\phi^\mathrm{(tot)}_\mathrm{cv}(\mathbf q,t)\right]}~,
\label{Dcv_par}
\end{equation}
and the total phase, $\phi^{(\mathrm{tot)}}_\mathrm{cv}$, is a sum of the dynamic and topological phases,
\begin{equation}
 \phi^{(\mathrm{tot)}}_\mathrm{cv}(\mathbf q,t)=\phi^\mathrm{(d)}_\mathrm{cv}(\mathbf q,t)+\phi^\mathrm{(T)}_\mathrm{cv}(\mathbf q,t)~.
 \label{phiQ}
 \end{equation}
Here, the topological phase is defined as $\phi^\mathrm{(T)}_\mathrm{cv}(\mathbf q,t)=\arg{\left[\mathcal{A}_\Vert(\mathbf q,t)\right]}$. This phase is 
the nontrival phase that the interband coupling amplitude acquires as a function 
of time. 
For two classes of trajectories, which correspond to points $\mathbf q$ outside and
inside of the separatrix, the topological phase behaves completely differently. 
This phase is displayed in Fig.\ \ref{fig:Phase_F_dot_D_CT_F0_5} for 
point $\mathbf q$ outside  (a) and inside (b) of the separatrix. The results are shown for $K^\prime$-valley. For the $K$-valley the corresponding phases have opposite signs. 
As we see, if point $\mathbf q$ is outside of the separatrix, the topological 
phase deceases with time near the $K^\prime$ point ($t\approx 0$) with 
the total change of $\approx -2\pi$. This total change is almost independent on the 
band gap, $2\Delta$.   
If point $\mathbf q$ is inside of the separatrix, the topological 
phase as a function of time increases near the $K^\prime$ point with 
the magnitude of the local increase that strongly depends
on the band gap. 
For zero band gap the topological phase remains constant, 
while with 
increasing of the band gap the magnitude of the local change of the 
topological phase near the $K^\prime$ point monotonically increases. 
Thus the topological phase for gapped graphene

As we see from Eq.~(\ref{Dcv_par}), the interband electron dynamics is determined by the total phase $\phi^{(\mathrm{tot)}}_\mathrm{cv}$,
which is a sum of the dynamic and topological phases. While the dynamic phase monotonically increases with time irrespective of 
the position of point $\mathbf q$ (inside or outside of the separatrix),
the dependence of the topological phase on time is different for 
points $\mathbf q$ inside and outside of the separatrix. As a result the interference of the dynamic and topological phases results in 
either a significant change of the total phase along the Bloch 
trajectory, which leads to small CB population, or mutual cancellation of the dynamic and topological phases, which results in coherent 
accumulation of the CB excitation amplitude and enhancement of CB
population. This is a topological resonance effect. 
For the right handed polarized pulse and for 
$K^\prime$ valley, see Fig.~\ref{fig:1CT_F0=0p5VpA_gapped_graphene}, the topological resonance occurs for
$\mathbf q$ points outside of the separatrix, while for 
$K$ valley the topological resonance occurs for $\mathbf q$ 
point outside of the separatrix. WIth increasing the band gap the topological resonance becomes more pronounced. 
Note that the conventional resonance can also be described as cancellation between the dynamic phase $2\Delta t/\hbar $ (where $2\Delta$ is excitation energy) and the field phase $-\omega t$, which occurs for $\omega\approx2\Delta/\hbar$.

For a case of left-handed pulse illustrated in Fig.\ \ref{fig:Phase_F_dot_D_CT_F0_5}, the topological resonance occurs for crystal momentum $\mathbf q$ inside the separatrix for the $K$-point and outside of the separatrix for the $K^\prime$-point.



\subsection{Linearly polarized pulse}

A unique feature of circularly polarized pulse is that an electron 
trajectory passes through a given point in the reciprocal space only once. As a result the topological resonance becomes well pronounced,
which manifests itself in large valley polarization and 
clear asymmetry between electron residual CB populations inside and outside 
of the separatrix. For linearly polarized pulse an electron 
passes through each given point in the reciprocal space twice.
In this case manifestation of the topological resonance in the 
residual CB population is suppressed, while the features of 
the topological resonance are visible during the pulse in both CB population and generated electric current.

Here we consider interaction of a linearly polarized pulse with 
gapped graphene. The pulse is polarized along $x$ axis and has the following profile
\begin{equation}
F_x=F_0(1-2u^2)e^{-u^2} ,
\end{equation}
where $F_0$ is the amplitude of the pulse, $u=t/\tau$ , and $\tau =1 $ fs. Similar to a circularly polarized pulse, we assume that initially 
the valence band is occupied and the conduction band is empty. 

\begin{figure}
\begin{center}\includegraphics[width=0.47\textwidth]{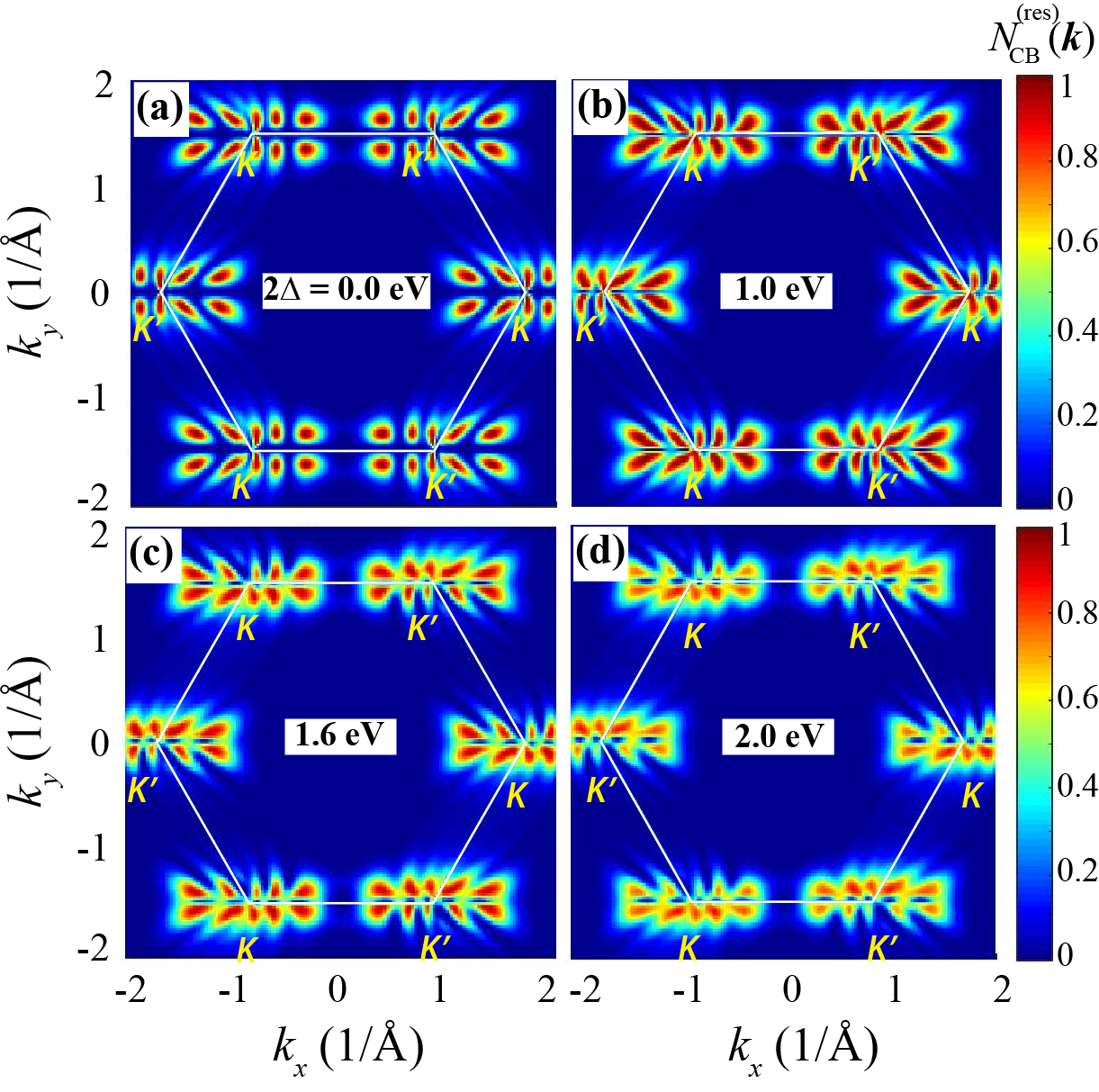}\end{center}
  \caption{(Color online) Residual CB population $N\mathrm{^{(res)}_\mathrm{CB}}(\mathbf{k})$ of gapped graphene with different bandgaps in the extended zone picture. The solid white line shows the first Brillouin zone boundary with $K, K^\prime$-points indicated. The amplitude of the optical field is 1 $\mathrm{V\AA^{-1}}$.
}
  \label{fig:Linear_Fx=1_gap_gapped_graphene}
\end{figure}

The residual CB population distribution in the reciprocal space is shown in Fig.~ \ref{fig:Linear_Fx=1_gap_gapped_graphene} for different values of the band gap. The hot spots are clearly visible in the 
population distribution. They are due to double passages by electrons
of the region near the $K$ ($K^\prime$) point during the pulse, 
which finally results in the corresponding interference pattern. 
Such hot spots were discussed in Ref.\  , where interaction of 
a linear optical pulse with graphene has been studied. For gapped 
graphene, the interference pattern becomes smeared, see 
Fig.~ \ref{fig:Linear_Fx=1_gap_gapped_graphene}, which is due to 
broadening of the interband dipole matrix element (non-Abelian 
Berry connection) for large band gaps. 

For gappless graphene, the CB population distribution is symmetric with respect to 
both $x$ and $y$-axes, see Fig. \ref{fig:Linear_Fx=1_gap_gapped_graphene}(a). For gapped graphene, 
the CB population distribution is centrosymmetric only without any axial 
symmetries. This is a manifestation of topological resonance for 
linearly polarized pulse. In this case, the 
population distribution is also chiral. Such chirality results in 
non-zero residual valley current in $y$ direction, while the $y$ 
component of the charge current vanishes after the pulse.

\begin{figure}
\begin{center}\includegraphics[width=0.47\textwidth]{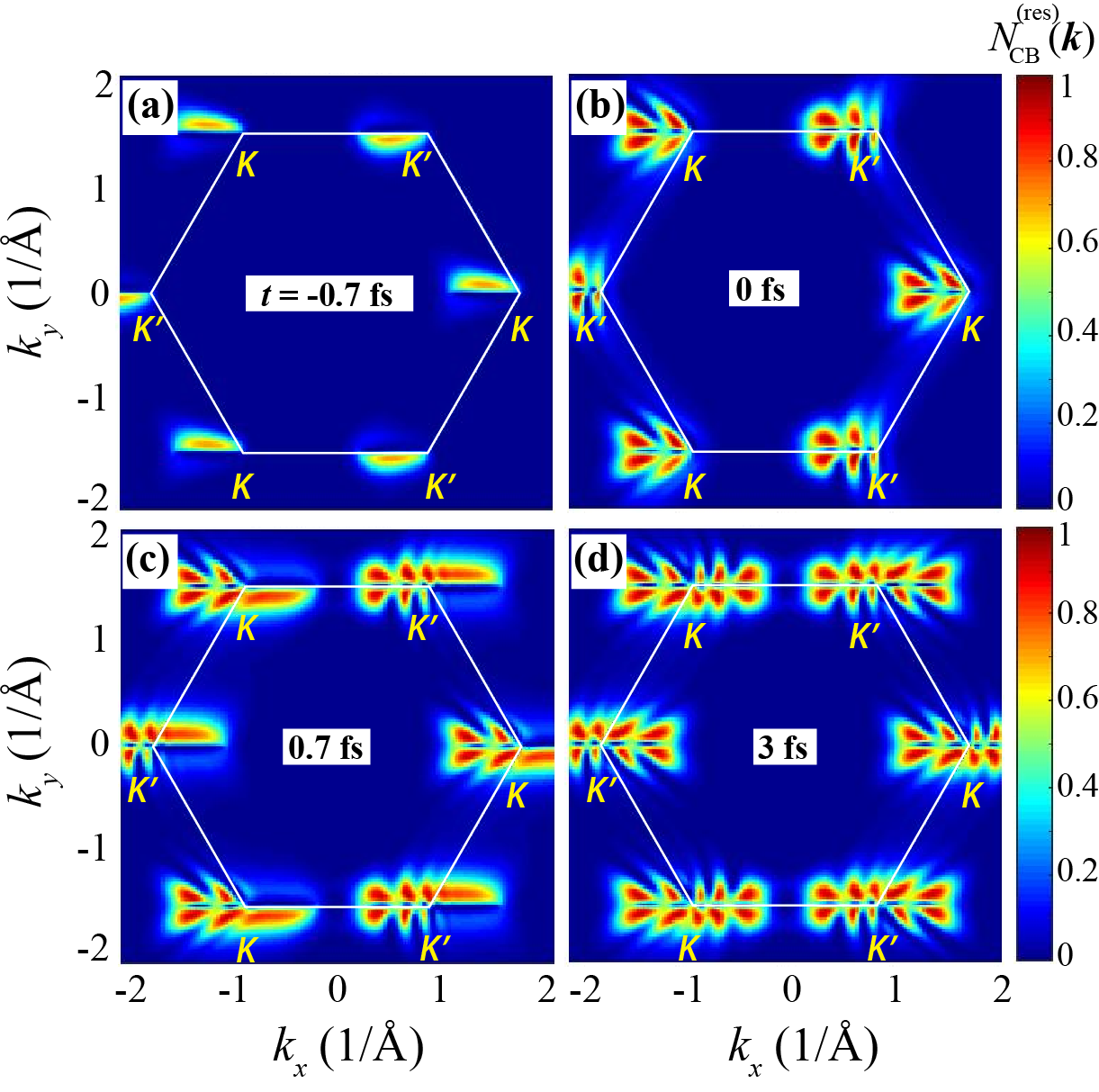}\end{center}
\caption{(Color online) CB population $N\mathrm{_\mathrm{CB}}(\mathbf{k})$  as a function of initial lattice vector for gapped graphene with bandgap 1.6 eV in the extended zone picture at different moment of time. The white solid line shows the first Brillouin zone boundary with $K, K^\prime$-points indicated. The applied pulse in linearly polarized in x direction and its amplitude is 1  $\mathrm{V\AA^{-1}}$.}
\label{fig:Linear_Fx=1_gap=1p6eV_gapped_graphene_time}
\end{figure}%

The topological resonance is more clearly visible in the time 
evolution of the CB population distribution, which is shown in 
Fig.~\ref{fig:Linear_Fx=1_gap=1p6eV_gapped_graphene_time} for gapped graphene with the band gap of 1.6 eV that is similar to the band gap of $\mathrm{MoS_2}$ monolayer. The amplitude of the pulse is 1 $\mathrm{V/\AA }$. At $t=-0.7$ fs, we see clear difference between CB populations 
near $K$ and $K^{\prime }$ points. This difference is due to topological resonance. Indeed, at $t<-0.7$ fs, electrons 
are moving to the 
right in the reciprocal space along $k_x$ axis and passing 
through the $K$ or $K^{\prime}$ point only once. Then the condition of 
the topological resonance is satisfied for $q$ points above the $K$ point 
and for $q$ points below the $K^{\prime }$ point. At $t=0$, 
electrons

The distribution of the CB population during the pulse is shown in Fig.  \ref{fig:Linear_Fx=1_gap=1p6eV_gapped_graphene_time} for gapped graphene with bandgap 1.6 eV (similar to $\mathrm{MoS_2}$ monolayer). The amplitude of the applied pulse is 1 $\mathrm{V\AA^{-1}}$. Here this distribution is not symmetric respect to x axis which is dramatically different from the case of graphene.
Initially, for $t\leq-0.7~\mathrm{fs}$, the applied field is negative so the electrons are accelerated to the right. As shown in Fig. \ref{fig:Linear_Fx=1_gap=1p6eV_gapped_graphene_time} (a) the left sides of valleys are populated sine the electrons located in this side cross the valley and due to the maximum of interband berry connection located at valleys the electrons are excited from Valence band to the conduction band. For $ -0.7 ~\mathrm{fs}<t\leq 0.7 ~\mathrm{fs}$, the field is positive which accelerate the electrons to the left and it creates interference fringes on the left side of valleys see Fig. \ref{fig:Linear_Fx=1_gap=1p6eV_gapped_graphene_time} (b) and populates the right side of the valleys see Fig. \ref{fig:Linear_Fx=1_gap=1p6eV_gapped_graphene_time} (c). For $0.7~ \mathrm{fs}<t\leq3~ \mathrm{fs}$, the field is negative which accelerates the electrons to the right and it creates interference fringes on the right side of valleys see Fig. \ref{fig:Linear_Fx=1_gap=1p6eV_gapped_graphene_time} (d). 

\begin{figure}
\begin{center}\includegraphics[width=0.47\textwidth]{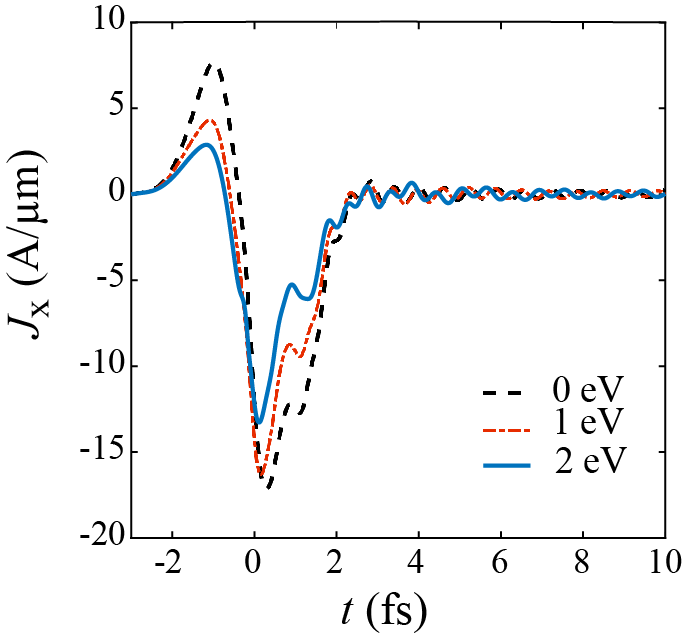}\end{center}
  \caption{Longitudinal current density, $J_\mathrm x$, in gapped graphene as a function of time for different bandgaps, 0 eV, 1 eV, and 2 eV. $F_\mathrm 0 = 1 \mathrm{V\AA^{-1}}$}
  \label{fig:Jx_total_F0=1}
\end{figure}

The applied linear polarized pulse generates a charge current in the direction of the pulse, for this case in the x-direction, we calculate the current using Eq. \ref{J} and it is shown in  Fig. \ref{fig:Jx_total_F0=1}. This current is generated due to the redistribution of electrons in CB and VB in the presence of the applied pulse. By increasing the bandgap the current decreases which can be understood by considering the total residual CB population where it decreases by increasing the bandgap see Fig. \ref{fig:Linear_Fx=1_gap_gapped_graphene}

\begin{figure}
\begin{center}\includegraphics[width=0.47\textwidth]{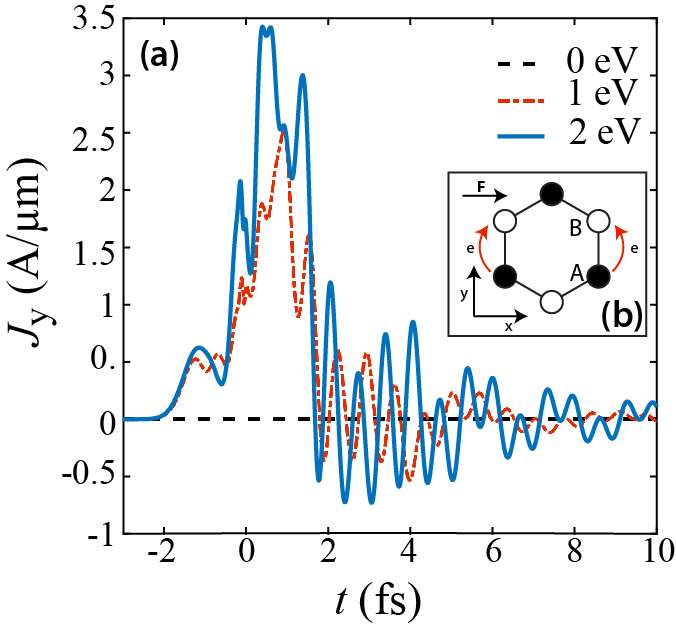}\end{center}
  \caption{(a) Hall current density, $J_\mathrm y$, in gapped graphene as a function of time for different bandgaps, 0 eV, 1 eV, and 2 eV. The amplitude of the applied field is $F_\mathrm 0 = 1 \mathrm{V\AA^{-1}}$. (b) The lattice structure of graphene with two sublattices, A and B, is shown here. Where A has higher on-site energy respect to B it causes the electron motion from A to B and creates Hall current in positive direction normal to the applied field.
  }
  \label{fig:Jy_total_F0=1}
\end{figure}

A striking finding here is that a photovoltaic Hall current is generated by the applied linear pulse in gapped graphene (see Fig. \ref{fig:Jy_total_F0=1})(a). By changing on-site energies to create a bandgap, one sublattice, A, gets higher on-site energy respect to the other sublattice, B, so electrons move from the sublattice with higher energy to the sublattice with lower energy see Fig. \ref{fig:Jy_total_F0=1} (b).  This Hall current is an addition to the longitudinal current generated in the direction of the applied pulse (see Fig. \ref{fig:Jx_total_F0=1}). Applying the pulse in opposite direction does not change the direction of the Hall current ($J_y$) which is determined by the on-site energies of sublattices A and B.
By increasing the bandgap modified by the on-site energies, the amplitude of the Hall current increases shown in from Fig. \ref{fig:Jy_total_F0=1}(a). This unbalanced current causes the net transferred charge which can be measured experimentally.

\begin{figure}
\begin{center}\includegraphics[width=0.47\textwidth]{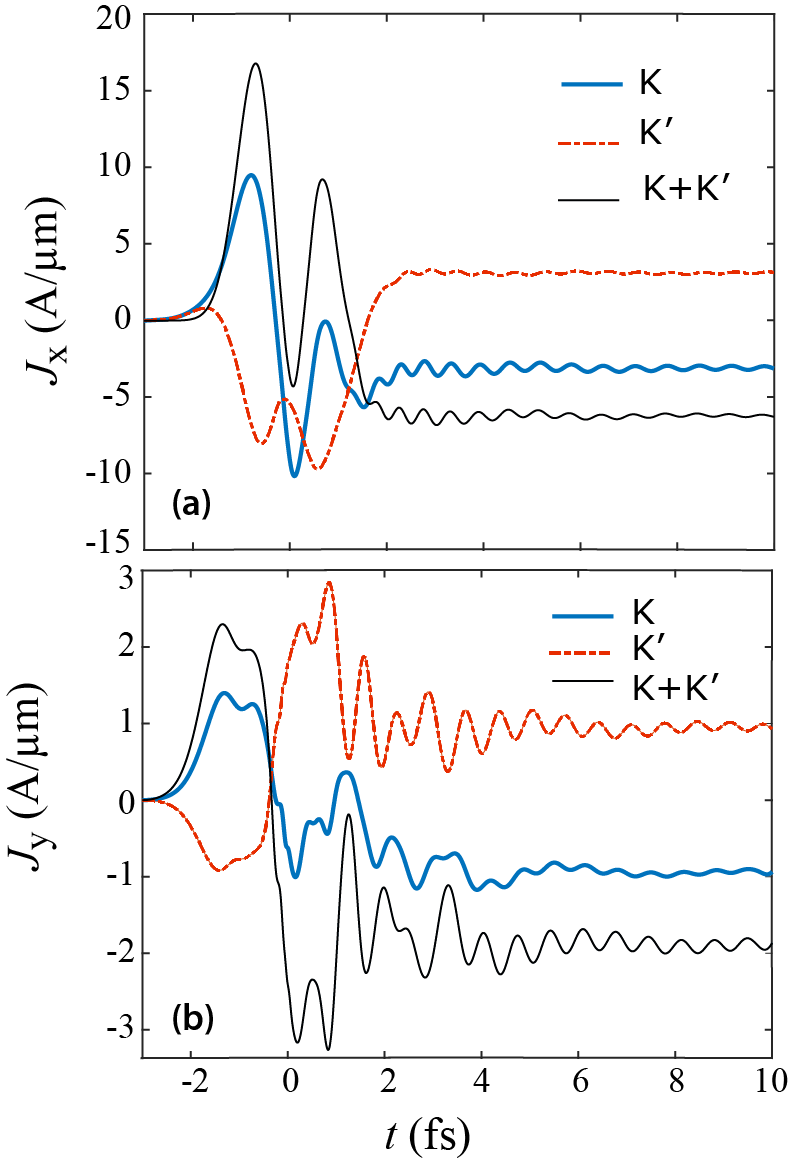}\end{center}
   \caption{Valley current density, (a) $J_\mathrm x$ and (b) $J_\mathrm y$ in gapped graphene (1 eV bandgap) as a function of time for $F_\mathrm 0 = 1 \mathrm{V\AA^{-1}}$}
  \label{fig:J_valley_F0=1_gap=1eV}
\end{figure}

In addition to the charge current and the Hall current, each valley generates a valley current shown in Fig.\ref{fig:J_valley_F0=1_gap=1eV} . Total valley current, $J^{(T)}$, characterized by the following expression 
\begin{equation}
J^{(T)}_\alpha=J^{(K)}_\alpha-J^{(K^ \prime)}_\alpha
\end{equation} 
where $\alpha$ shows the direction of the current.


\section{conclusion}
We demonstrated that a fundamentally fastest valley polarization could be induced in gapped graphene by a single oscillation circularly polarized pulse. This effect is similar to TMDC where the circular pulse populates one valley significantly respect to the other depending on the polarization of the pulse. We also showed the effect of bandgap on the valley polarization. The existence of the bandgap is necessary to have a valley polarization since it causes a gradual accumulation of the topological phase along the Bloch $\mathbf k-$space electron trajectory, which is necessary to compensate
the gradually accumulating dynamic phase.

 Also, we predicted that the distribution of the CB population in the reciprocal space induced by the applied linear pulse is chiral. This electron distribution can be observed by time resolve angle-resolved photoelectron spectroscopy (tr-ARPES). The linear pulse generates a longitudinal current in the direction of the field and a photovoltaic Hall current normal to the field in gapped graphene. This Hall current is generated in the absence of a magnetic field by a linearly polarized pulse. While applying pulse in the opposite direction changes the direction of the longitudinal current it does not change the direction of the Hall current which is only affected by the on-site energies of different sublattices. Additionally, the unbalanced profiles of currents generate transferred charges in the direction of the applied field and the direction normal to the applied field.
 In addition to the charge current, there is a nonzero net valley current which can be measured experimentally. 
 
 The predicted ultrafast valley polarization has the potential to be used in ultrafast quantum memory devices for quantum information processing.
 
%

\end{document}